\documentstyle[eqsecnum,aps,prd,epsf]{revtex}
\newcommand{\para}{/\!\!/}  
\newcommand{\be}{\begin{equation}}  
\newcommand{\ee}{\end{equation}}  
\newcommand{\bea}{\begin{eqnarray}}  
\newcommand{\eea}{\end{eqnarray}}  
\newcommand{\del}{\delta}  
   
\newcommand{\non}{\nonumber}  
\newcommand{\tD}{{}^{(3)}\!D}
\begin{document}  
%----------------------------------------------------------------------------  
%\begin{titlepage}

%%%    
\def\abstract#1
{\begin{center}{\large Abstract}\end{center}\par #1}
\def\title#1{\begin{center}{#1}\end{center}}
\def\author#1{\begin{center}{#1}\end{center}}
\def\address#1{\begin{center}{\it #1}\end{center}}
%%%   

%----------------------------------------------------------------------------  
\def\pubnum{387/COSMO-88} 

%%%
\hfill
\parbox{6cm}{{TIT/HEP-\pubnum} \par 
%\today
}   
\par 
%%% 
\vspace{7mm}

%----------------------------------------------------------------------------  
\title{ 
{\LARGE {\bf
        Equation of motion for a domain wall \\ 
        coupled to gravitational field 
      }}
       }
\vskip 8mm 
\author{ {\large  
        Akihiro Ishibashi\footnote{E-mail:akihiro@th.phys.titech.ac.jp}
        and 
        Hideki Ishihara\footnote{E-mail:ishihara@th.phys.titech.ac.jp}
         }
       }
\vskip 2mm 
\address{ {\large {\it 
        Department of Physics, \\ Tokyo Institute of Technology, \\ 
        Oh-Okayama Meguro-ku, Tokyo 152, Japan
                  }
          }
        }

%----------------------------------------------------------------------------- 

%----------------------------------------------------------------------------- 
\vskip 10mm     
\abstract{  
\noindent 
         The equation of motion for a domain wall coupled to gravitational 
         field is derived from the Nambu-Goto action. 
         The domain wall is treated as a source of gravitational field. 
         The perturbed equation is also obtained with gravitational back 
         reaction on the wall motion taken into account. 
         For general spherically symmetric background case, the equation 
         is expressed in terms of the gauge-invariant variables. 
} 
\vskip 3mm

\noindent 
PACS number(s): 04.30.Nk, 04.40.Nr, 98.80.Cq, 98.80.Hw
% 

%\maketitle

%----------------------------------------------------------------------------- 
\section{Introduction}  
%----------------------------------------------------------------------------- 

As a possible evidence of cosmological phase transitions in the early 
universe, topological defects may remain somewhere in our universe. 
The physics of spacetimes containing the defects such as cosmic strings 
and domain walls has been investigated 
extensively~(see, e.g., Ref.~\cite{VS}). 
In particular, domain walls as the boundary of a vacuum bubble in a 
false vacuum sea play an essential role in the open inflationary 
cosmology~\cite{Oboi} recently developed. 
There have been increasing interests in dynamics of domain walls 
in cosmology. 

Since, in most situations of cosmological interest, 
the thickness of domain walls can be neglected compared to all the other 
scales we shall consider infinitely thin domain walls in what follows. 
In this thin wall approximation, the trajectory of a thin domain wall 
in a spacetime is described as a hypersurface called the {\em world sheet}. 
In studies of dynamics of domain walls, 
especially in perturbation theory, we often assume 
that gravitational effects of domain walls are negligible. 
The dynamics of the wall is governed by the Nambu-Goto action 
on a fixed background spacetime. 
We also expect that the equation derived from the Nambu-Goto 
action could well describe the behavior of the wall even 
in the case that gravitational back reaction on the wall motion 
is taken into account. 

In perturbation theory, the small deformation of a domain wall is described 
by a single scalar field $\phi$ living in the background world 
sheet~\cite{GV,G}. On the assumption neglecting gravitational effects 
of the domain wall, the perturbation $\phi$ obeys the Klein-Gordon equation 
in the form 
\be 
   \left( \Box_3 - m^2 \right) \phi = 0, 
\label{eq:KG}
\ee 
where the mass term $m^2$ is expressed in terms of the difference of 
the vacuum energy densities on the two sides of the wall, 
the surface energy density, the three-scalar curvature of the unperturbed 
world sheet, and the spacetime Ricci tensor at the world sheet. 
The scalar field $\phi$ couples to the exterior spacetime geometry 
only through the mass term. 
Since this equation has oscillatory solutions we naively expect that 
the oscillation of $\phi$ might generate gravitational waves via the perturbed 
Einstein equations and would be damped gradually 
by the emission of the gravitational waves. 

On the other hand, we can see qualitatively different behavior of 
domain walls, using the metric junction formalism of Israel~\cite{Is}. 
The spacetime containing a thin domain wall as a source of gravitational 
field becomes singular at the wall hypersurface. 
The metric junction formalism is one of the ways to treat 
such singular hypersurfaces. 
Since the junction condition is derived from the Einstein equations, 
the influence of gravitational field on the wall motion is automatically 
taken into account in this formalism. 
The perturbations of domain walls interacting with gravitational waves 
are investigated by using the formalism by Kodama, et.al.~\cite{KIF} 
in a locally flat background case and by the present authors~\cite{II} 
in the one-bubble open inflation~\cite{Oboi} background case. 
In these works, thanks to the high symmetry ${\rm O(3,1)}$ of the background 
geometry, the perturbed motion of the wall coupled to gravitational waves 
can be solved explicitly. 
It turns out that, once the gravitational back reaction is 
taken into account, the oscillatory behavior of 
the deformation of the domain wall disappears. 
Furthermore, the domain wall loses its dynamical degree of freedom 
and the perturbed motion of the wall is completely accompanied 
with gravitational perturbations. 
It is remarkable that the results indicate that the gravitational 
back reaction cannot be ignored even in the first order of 
the perturbation amplitude.  

In general, when we resolve the coupled system of a domain wall and 
gravitational perturbations, our task is to solve 
the perturbed Einstein equations and the equation of motion (EOM) for 
the domain wall coupled to gravitational field simultaneously. 
Indeed this has been accomplished in the works~\cite{KIF,II}, 
but the relation between the wall behavior in these analyses 
and the EOM for the domain wall derived from the Nambu-Goto action remains 
being unclear.

As is well known, the EOM for a source of gravitational field is built into 
the Einstein equations via the Bianchi identity. 
Thus the EOM for a domain wall coupled to gravitational field 
should be consistent with the Einstein equations. 
Furthermore, as mentioned above, the world sheet of a thin domain wall 
as a concentrated source is singular hypersurface, 
hence the dynamics of the wall is that of the singularity. 
Then there naturally arises a non-trivial question whether the motion of 
such a singular source is well described by the equation 
derived from the Nambu-Goto action on a fixed background spacetime. 

In this paper, we shall derive the EOM for domain walls coupled to 
gravitational fields from an action principle. 
We explicitly show the perturbed EOM for a domain wall 
coupled to gravitational perturbations in generic background. 
When we treat perturbations of gravitational fields, we should 
give care to gauge modes. 
For general spherically symmetric background case, as a simple example, 
we express the EOM in terms of the gauge-invariant perturbation variables. 
We can see that the EOM obtained here actually coincides with 
that obtained by the metric junction formalism in Ref.~\cite{II} 
in the case of a vacuum bubble nucleation in de Sitter spacetime.

In the next section, we shall develop the procedure to obtain EOM for 
a domain wall coupled to gravitational field from the Nambu-Goto action 
to be consistent with the Einstein equations. 
In Sec.~\ref{sec:pertEq}, we first obtain the perturbed EOM for a domain wall 
coupled to gravitational perturbations without imposing any symmetry on 
the background geometry. Next we introduce the gauge-invariant perturbation 
variables for the general spherically symmetric background case 
and represent the perturbed equation explicitly in terms of them. 

Throughout the paper, the unit $c=1$ is adopted and $8\pi G$ is denoted 
by $\kappa$, the signature of the spacetime metric is taken to be $(-,+,+,+)$. 
The definitions of geometrical quantities 
and the notation are essentially the same as those in Ref.~\cite{II}. 

%----------------------------------------------------------------------------- 
\section{General formula} 
\label{sec:gefo}
%----------------------------------------------------------------------------- 

Consider a four-dimensional spacetime $(M,g_{\mu \nu})$ 
with a continuous metric containing a domain wall whose world sheet 
$\Sigma$ divides the spacetime manifold $M$ into two regions: 
$M_-$ and $M_+$. 
Provided that the whole spacetime manifold $ M = M_{+} \cup \Sigma \cup M_{-}$ 
is smooth and $\Sigma$ is also smooth as a submanifold in $M$, 
a smooth coordinate system $\{x^\mu \}$ (the indices $\mu, \nu ...$ 
run over 0,1,2,3) is taken in a neighborhood of $\Sigma$ in $M$. 
Then the embedding of $\Sigma$ in $M$ is described by 
\be
        x^\mu = x^\mu (\zeta^i), 
\ee 
where $\{ \zeta^{i} \}$ (the indices $i,j, ...$ run over 0,1,2) is 
an intrinsic three-coordinate system assigned on $\Sigma$. 
The unit normal vector $n^{\mu}$ of $\Sigma$ 
pointing toward $M_+$ and the tangent vectors defined as 
\be 
e^{\mu}_{i} 
         := \frac{\partial x^{\mu}}{\partial \zeta^{i}}, 
\ee 
are characterized by
\be  
        g_{\mu \nu}n^{\mu}n^{\nu} = 1 , \qquad 
        g_{\mu \nu} e^{\mu}_{i}n^{\nu} = 0. 
\ee  
Then, the intrinsic metric $q_{ij}$ on $\Sigma$ is induced as   
\bea
        q_{ij} := g_{\mu \nu}e^{\mu}_{i}e^{\nu}_{j}, 
\label{qij} 
\eea  
and its inverse satisfies  
\be 
        q^{ij} e^{\mu}_{i}e^{\nu}_{j} = g^{\mu \nu} - n^{\mu}n^{\nu}. 
\ee  

The action for the coupled system of a domain wall and gravitational field 
is given by 
%-----------------------------------------------------------------------------
\be
   S = - \sigma \int_{\Sigma} d^3\zeta \sqrt{-q} 
       - \rho_{+} \! \int_{M_{+}}d^4 x \sqrt{-g} 
       - \rho_{-} \! \int_{M_{-}}d^4 x \sqrt{-g} 
       + \frac{1}{2 \kappa}\int d^4 x \sqrt{- g} R . 
\label{WGaction} 
\ee  
The first term is so-called the Nambu-Goto action for the domain wall 
which is in proportion to the proper three-volume of wall's 
world sheet. The constant $\sigma$ represents 
the surface energy density of the wall. 
The second and the third terms are proportional to the four-volume of 
the $M_+$ and $M_-$ regions respectively. 
The constants $\rho_\pm$ are the vacuum energy densities of 
the $M_\pm$ regions respectively. 
The last term is the Einstein action, 
where $ R $ is the scalar curvature associated with $ g_{\mu \nu}$. 
%-----------------------------------------------------------------------------

First, consider the variation of the action~(\ref{WGaction}) with respect 
to small changes in $x^{\mu}(\zeta^i)$ on $\Sigma$, 
\be 
   x^{\mu} \longrightarrow x^{\mu} + \breve{\delta} x^{\mu}. 
\label{variation} 
\ee 
Then we obtain the equation~\cite{GV,G}    
\be         
     \Box_3 x^{\mu} + {\mit \Gamma}^{\mu}{}_{\nu \lambda} q^{ij} 
     e^{\nu}_{i}e^{\lambda}_{j} + \frac{\rho}{\sigma} n^{\mu} = 0, 
\label{eq:eomow}
\ee  
where $\Box_3$ is the d'Alembertian in $\Sigma$ defined by 
\be
\Box_3 x^{\mu}
      := \frac{1}{\sqrt{-q}}\partial_i (\sqrt{-q} q^{ij}\partial_j x^{\mu}) 
      = {}^{(3)}D{}^{j} e^{\mu}_{j}, 
\ee
${\mit \Gamma}^{\mu}{}_{\lambda \nu}$ is the Levi-Civita connection 
associated with $g_{\mu \nu}$, ${}^{(3)}D{}^{j}$ is 
the covariant derivative with respect to $q_{ij}$, 
and $\rho := \rho_+ - \rho_-$. 

The tangential directions of $\breve{\del} x^{\mu}$ to $\Sigma$ 
correspond to diffeomorphisms, 
$ \zeta^i \rightarrow \zeta^i + \breve{\del} \zeta^i $, 
on $\Sigma$ and the transverse direction is the only degree 
of freedom for physical motion. Hence physically relevant component 
of Eq.~(\ref{eq:eomow}) is the normal one: 
\be         
   n_\mu \Box_3 x^{\mu} + n_\mu {\mit \Gamma}^{\mu}{}_{\nu \lambda} q^{ij} 
     e^{\nu}_{i}e^{\lambda}_{j} + \frac{\rho}{\sigma} = 0. 
\label{eq:Nor.eomow}
\ee  
By the Gauss-Weingarten equation 
\be
{}^{(3)}D_{i} e^{\mu}_{j} + {\mit \Gamma}^{\mu}{}_{\nu \lambda}
                                  e^{\nu}_{i} e^{\lambda}_{j} 
            = n^{\mu} K_{ij}, 
\label{eq:GW}
\ee  
Eq.~(\ref{eq:eomow}) is further simplified in the form
\be
  K = - \frac{\rho}{\sigma}. 
\label{eq:WMpm}
\ee 
Here, $K$ is the trace of the extrinsic curvature defined by 
\bea 
    K_{ij} := - e^{\mu}_{i}e^{\nu}_{j}\nabla_{\nu}n_{\mu}. 
\eea 

%-----------------------------------------------------------------------------

Next, taking the variation of the action in $g^{\mu \nu}$, we obtain 
the Einstein equations 
\be 
   R_{\mu \nu} - \frac{1}{2} R g_{\mu \nu} = \kappa T_{\mu \nu}. 
\label{eq:Einstein}
\ee 
For convenience, let us introduce 
the Gaussian normal coordinate system $\{\chi, \zeta^i\}$ in a neighborhood 
of $ \Sigma $ in which the spacetime metric takes the form 
\be 
        g_{\mu \nu} dx^{\mu}dx^{\nu} 
                = d \chi^2 + q_{ij} d\zeta^{i} d\zeta^{j}. 
\label{GaussNormal}  
\ee 
Without loss of generality, one may choose $ \chi = 0 $ 
on $\Sigma$ and $ \chi > 0 (< 0)$ on $M_{+}(M_{-}) $, respectively. 
In this coordinate system the energy-momentum tensor is expressed as 
\be 
   T_{\mu \nu} 
         = - \sigma q_{\mu \nu}(\zeta^i) \delta (\chi) 
                - \rho g_{\mu \nu} \theta(\chi), 
\label{def:emtensor}
\ee 
where $q_{\mu \nu}$ is the four-dimensional extension of $q_{ij}$, 
$\delta(\chi)$ the $\delta$-function, and $\theta(\chi)$ the step function. 

%-----------------------------------------------------------------------------
It should be noted that the whole spacetime 
manifold $M$ can be regarded as a hybrid of $M_+$ and $M_-$. 
The boundary hypersurfaces 
$\partial M_{\pm}$ should be matched  with $\Sigma$. 
Since the coordinates $x^\mu$ are smooth in a neighborhood of 
$ \Sigma $ the tangent vectors $e^{\mu}_{i}$ are continuous at $ \Sigma $. 
Then the intrinsic metric $q_{ij}$ defined by Eq.~(\ref{qij}) is continuous 
and well defined at $\Sigma$. 
The three-Ricci tensor ${}^{(3)}R^i{}_j$ in $\Sigma$ is also continuous. 

On the other hand, since Eq.~(\ref{def:emtensor}) has a singular 
term which is proportional to the $\delta$-function 
and the Einstein equations~(\ref{eq:Einstein}) contain 
the second derivatives of the spacetime metric, 
the first derivatives of the spacetime metric become
in general discontinuous at $\Sigma$, i.e., 
$(\partial_\chi g_{ij})|_{-} 
\neq (\partial_\chi g_{ij})|_{+}$. 
The related quantities such as the extrinsic curvature 
$K_{ij}$ is not well defined on $\Sigma$. Thus Eqs.~(\ref{eq:Nor.eomow}) 
and (\ref{eq:WMpm}) have ambiguities in the evaluation 
of ${\mit \Gamma}^{\mu}{}_{\lambda \nu}$ and $K_{ij}$ at $\Sigma$.

Let us see how to evaluate the discontinuous quantity $K{}^i{}_j$ 
on $\Sigma$ from the view point of consistency with the Einstein equations 
on the whole spacetime~$ (M, g_{\mu \nu}) $. 
Let $\Sigma_\chi$ be a $\chi=const.$ hypersurface 
($\Sigma = \Sigma_{\chi=0}$). In the Gaussian normal coordinate system, 
the unit normal vector to $\Sigma_\chi$ and 
the extrinsic curvature of $\Sigma_\chi$ , which is well defined 
if $\Sigma_\chi \neq \Sigma$, are represented respectively by 
\bea 
        n^{\mu} &=& (\partial_{\chi})^{\mu}, 
\\ 
K_{ij} 
     &=& - \frac{1}{2} \partial_{\chi} g_{ij}. 
\label{Kij}
\eea 
The $(i,j)$ components of the Einstein equations are formally expressed as 
\bea  
G{}^i{}_j 
   &\equiv&  
       {}^{(3)}\!R^i{}_j + \frac{1}{2} {}^{(3)}\!R \delta^i{}_j
       - K K^i{}_j 
       - \frac{1}{2} \delta^i{}_j 
                 \left( 
                       2 q^{lm}{}^{(3)}\!R_{lm} - K^{lm} K_{lm} + K^2 
                 \right) 
       + \partial_\chi K^i{}_j 
\non \\ 
    &=& 
       - \kappa \del^i{}_j \sigma 
                        \left\{
                               \del(\chi) + \frac{\rho}{\sigma } \theta(\chi)
                        \right\}.
\label{Ein:Gij}
\eea  
Then the extrinsic curvature of $\Sigma$ should be given by 
\bea 
    K{}^i{}_j|_\Sigma 
    &:=& 
        \lim_{\epsilon \rightarrow 0} 
            \int^{\epsilon}_{- \epsilon} d\chi 
                                              K{}^i{}_j \del (\chi) 
\non \\ 
    &=& 
       - \frac{1}{2 \kappa \sigma} 
        \lim_{\epsilon \rightarrow 0} 
            \int^{\epsilon}_{- \epsilon} d\chi 
        \left(  
               K{}^i{}_m \partial_\chi K{}^m{}_j 
               + K{}^m{}_j \partial_\chi K{}^i{}_m
        \right)  
\non \\ 
    &=&  
       - \frac{1}{2 \kappa \sigma} 
        \lim_{\epsilon \rightarrow 0} 
            \int^{\epsilon}_{- \epsilon} d\chi 
        \partial_\chi
        \left(  
               K{}^i{}_m K{}^m{}_j 
        \right)  
\non \\ 
    &=&
       - \frac{1}{2 \kappa \sigma} 
         \lim_{\epsilon \rightarrow 0} 
       \left( 
             \overline{K}{}^i{}_m
        \int^{\epsilon}_{- \epsilon} d\chi 
          \partial_\chi K{}^m{}_j
       +     
        \overline{K}{}^m{}_j
        \int^{\epsilon}_{- \epsilon} d\chi 
          \partial_\chi K{}^i{}_m 
       \right) 
\non \\
    &=&  
       \frac{1}{2} 
       \lim_{\epsilon \rightarrow 0} 
       \left\{  
              \overline{K}{}^i{}_m
        \int^{\epsilon}_{- \epsilon} d\chi 
             \del{}^m{}_j \del (\chi)
       +     
       \overline{K}{}^m{}_j
        \int^{\epsilon}_{- \epsilon} d\chi 
             \del{}^i{}_m \del (\chi) 
       \right\}
\non \\ 
    &=& 
       \overline{K}{}^i{}_j, 
\label{def:KonSigma} 
\eea 
where Eq.~(\ref{Ein:Gij}) is used for the second and the fifth lines, 
and $\overline{K}{}^i{}_j$ is defined as 
\bea  
    \overline{K}{}^i{}_j
            := \frac{1}{2}
               \lim_{\epsilon \to 0} 
                        \left\{ 
                                K{}^i{}_j|_{+ \epsilon} 
                                + K{}^i{}_j|_{- \epsilon} 
                        \right\} 
            = \frac{1}{2}
                        \left\{ 
                                K{}^{i+}_{~j}
                                + K{}^{i-}_{~j}
                        \right\}. 
\label{Kave}
\eea 
The extrinsic curvature $K_{ij}^+ (K_{ij}^-)$ of 
the boundary hypersurface $\partial M_+ (\partial M_-)$ 
is defined by the right (left) differential coefficient 
$\partial_\chi g_{ij}|_+ (\partial_\chi g_{ij}|_-)$ 
on the hypersurface. 
We should replace the extrinsic quantities on $\Sigma$ 
by the algebraically averaged one. 
Hereafter, we use \lq average\rq\ in the sense of Eq.~(\ref{Kave}).  

We therefore should recognize the extrinsic quantities which appear in 
Eq.~(\ref{eq:Nor.eomow}) (and equivalently Eq.~(\ref{eq:WMpm})) 
as the averaged one.
Thus Eq.~(\ref{eq:Nor.eomow}) should be replaced by 
\be 
   {n}_{\mu}\Box_3 x^{\mu} 
     + {n}_{\mu} \overline{{\mit \Gamma}}^{\mu}{}_{\nu \lambda} q^{ij} 
       e^{\nu}_{i}e^{\lambda}_{j} 
                    + \frac{\rho}{\sigma}  = 0. 
\label{Eom:wmg}
\ee   
This expression is independent of coordinate choice. 
Equivalently, Eq.~(\ref{eq:WMpm}) should be 
\be
  \overline{K} = - \frac{\rho}{\sigma}. 
\label{eq:WM} 
\ee

We note that, in fact, we can see that Eqs.~$K_\pm = \rho/\sigma$ 
are incompatible with the continuity of the intrinsic metric $q_{ij}$ 
from the junction conditions~\cite{KIF,II,HS,BKT,BGG}.  
For example, in the case of a vacuum bubble~\cite{II}, 
if $K_\pm = \rho/\sigma$ expressed the bubble wall motion, 
it would imply $r^2_- \neq r^2_+$ at $\Sigma$ 
from Eq.~(2.44) in Ref.~\cite{II}, 
though $r^2_\pm$ are the components of the intrinsic metrics $(q_{ij})_\pm$. 

%----------------------------------------------------------------------------- 
On the other hand Eq.~(\ref{Eom:wmg}) (and equivalently Eq.~(\ref{eq:WM})) 
describes the embedding of $\Sigma$ into the physical spacetime 
$(M,g_{\mu \nu})$ being consistent with the Einstein equations 
and the motion of the domain wall.

%----------------------------------------------------------------------------  

Here, we briefly summarize the procedure how to derive the EOM for 
a domain wall coupled to gravitational field. 
(i) 
Consider the half spacetimes $M_+$ and $M_-$ separately and 
take the variations of the actions~(\ref{WGaction}) in the extrinsic 
coordinates $x^\mu$ of wall and get tentative equations of motion. 
(ii) 
Take an average of these two equations, 
so we get the desired equation for $\Sigma$ in $(M, g_{\mu \nu})$.

%----------------------------------------------------------------------------  
\section{Perturbed equation } 
\label{sec:pertEq}
%----------------------------------------------------------------------------  

Let us consider the perturbed EOM for a domain wall. 
Following the procedure proposed in the previous section, 
we first take the perturbations of the tentative 
equation~(\ref{eq:Nor.eomow}) formally, and then replace 
the extrinsic quantities on $\Sigma$ by the averaged ones.

\subsection{Perturbed EOMs on generic background spacetimes}
%----------------------------------------------------------------------------  
We concentrate on $M_+$ and suppress `+' to keep the simplicity. 
Let $\tilde{\del}$ be a perturbation of the extrinsic coordinates 
$x^{\mu}$ on the wall:
\be
\tilde{x}^{\mu} := x^{\mu} + \tilde{\del} x^{\mu}, 
\label{prtrb:phys}
\ee 
and consider the perturbation of Eq.~(\ref{eq:Nor.eomow}). 
Note that the perturbation $\tilde{\del}$ in Eq.~(\ref{prtrb:phys}), 
which is formally the same as the mathematical variation $\breve{\delta}$ 
in Eq.~(\ref{variation}), is a physical perturbation. 
Since the transverse motion is physically relevant, 
the wall displacement $\tilde{\del} x^{\mu}$ can be described with 
a single scalar field $\phi(\zeta^i)$ in $\Sigma$ by 
\be 
\tilde{\del} x^{\mu} = \phi n^{\mu}. 
\label{scalar}
\ee  

In Refs.~\cite{GV,G}, it is considered that the perturbations of the wall 
intrinsic metric are induced by $\tilde{\del}x^{\mu}$ 
through the relation $q_{ij}(x) = g_{\mu \nu}(x)e^{\mu}_{i}e^{\nu}_{j}$. 
However in the coupled system of a domain wall and gravitational field, 
the wall motion also generates metric perturbations. 
Then we should take account of them, described 
by $\del'$ as
\be
        \del' : g_{\mu \nu}(x) \longrightarrow g'_{\mu \nu}(x) 
                        := g_{\mu \nu}(x) + h_{\mu \nu}(x). 
\ee 

Now, full perturbation is $\del := \tilde{\del} + \del'$. 
The spacetime metric then transforms as  
\bea  
\del : g_{\mu \nu}(x) \longrightarrow g'_{\mu \nu}(\tilde{x}) 
        &=&{} 
              g_{\mu \nu}(x + \tilde{\del}x) + h_{\mu \nu} (x) 
\non \\  
   &=& {} 
              g_{\mu \nu}(x) - 2\phi K_{\mu \nu}(x) + h_{\mu \nu}(x), 
\label{del:g}
\eea
where $K_{\mu \nu}$ is the four-dimensional extension of $K_{ij}$ 
of the background world sheet. 
The wall intrinsic metric $q_{ij}$ transforms as 
\bea
\del : q_{ij}(x) \longrightarrow q'_{ij}(\tilde{x}) 
  &=&{} 
       \left\{ 
               g_{\mu \nu}(\tilde{x}) + h_{\mu \nu}(\tilde{x}) 
       \right\} 
                {e}^{\mu}_{i} (\tilde{x}) {e}^{\nu}_{j} (\tilde{x})
\non  \\  
  &\simeq&{}  
          q_{ij}(x) - 2\phi K_{ij}(x) + h_{ij}(x), 
\eea 
where 
\be 
 {e}^{\mu}_{i} (\tilde{x}) 
              := \frac{\partial \tilde{x}^{\mu}}{\partial \zeta^{i}} 
              = e^{\mu}_{i} (x) 
                + n^{\mu} \frac{\partial \phi}{\partial \zeta^{i}} , 
\ee  
and we use the orthogonality $e^{\mu}_{i}n_{\mu} = 0$. 
Thus, 
\bea
&&{} 
    \del q_{ij}(x) = - 2\phi K_{ij}(x) + h_{ij}(x), 
\label{del:q}
\\ 
&&{} 
    \del q^{ij}(x) := - q^{il}q^{jk} \del q_{lk}(x) 
                    =  2\phi K^{ij}(x) - h^{ij}(x). 
\label{del:q-1}
\eea  

From Eq.~(\ref{eq:eomow}), the perturbation of Eq.~(\ref{eq:Nor.eomow}) 
reduces to 
\be
  n_{\mu}
    \del 
     \left( 
           \Box_3 x^{\mu} 
            + {\mit \Gamma}^{\mu}{}_{\nu \lambda} 
              q^{ij}e^{\nu}_{i} e^{\lambda}_{j} 
            + \frac{\rho}{\sigma} n^{\mu} 
      \right) = 0.  
\label{eq:normaldir}
\ee 
The first term in the LHS of Eq.~(\ref{eq:normaldir}) is 
\bea  
&&{} 
  n_{\mu} \del 
            \left(
                   \Box_3 x^{\mu}
            \right)
  =   n_{\mu} {}^{(3)}\!D^{j} \del e^{\mu}_{j}
  =  \Box_3 \phi, 
\eea 
where we use the relation 
$ 
\del e^{\mu}_{j} 
   = {\partial \del x^{\mu}}/{\partial \zeta^j} 
   = n^{\mu} \partial_{j} \phi 
$. 
After some calculation, which is shown in Appendix~\ref{app:cal}, 
the second term in the LHS of Eq.~(\ref{eq:normaldir}) becomes
\bea 
&&{} n_{\mu} \del{}
      \left(
            {\mit \Gamma}^{\mu}{}_{\nu \lambda} 
                              q^{ij} e^{\nu}_{i}e^{\lambda}_{j}
      \right) 
=
  {}^{(3)}\! D^j h_{\chi j} - \frac{1}{2} q^{ij} \partial_{\chi} h_{ij} 
   - K h_{\chi \chi} 
   - K_{ij} h^{ij} 
   + 2 \left(
             K_{ij} K^{ij} + \frac{1}{2} q^{ij} \partial_{\chi} K_{ij} 
       \right) \phi. 
\label{del:Gamma}
\eea 
It should be noted that $K_{ij}$ and $\partial_{\chi} K_{ij}$ 
in Eq. (\ref{del:Gamma}) 
are defined  by the right differential coefficient in $M_+$ 
on $\partial M_+$ 
From $ g_{\mu \nu} n^{\mu} n^{\nu} = 1$, 
the third term in the LHS of Eq.~(\ref{eq:normaldir}) becomes 
\bea 
&&{} 
    n_{\mu} \del \left(
                       \frac{\rho}{\sigma} n^{\mu}
                 \right)
    = - \frac{1}{2} \frac{\rho}{\sigma} h_{\chi \chi}. 
\eea 
We therefore find that Eq.~(\ref{eq:normaldir}) reduces to 
\bea 
&&{}
 \left(
         \Box_3 + 
                2 K_{ij} K^{ij} + q^{ij} \partial_{\chi} K_{ij} 
 \right) \phi (\zeta^i) 
\non \\ 
&&{} 
\qquad \qquad \qquad \qquad 
 + {}^{(3)}\!D^j h_{\chi j} - \frac{1}{2} q^{ij} \partial_{\chi} h_{ij} 
   - K_{ij} h^{ij} 
 - \left(
         K + \frac{1}{2} \frac{\rho}{\sigma} 
   \right) h_{\chi \chi} 
 = 0. 
\eea 
With the aid of the relation 
\be 
   - \frac{1}{2} q^{ij} \partial_{\chi} h_{ij} - K_{ij} h^{ij} 
   =   \frac{1}{2} \partial_{\chi} h^j{}_j, 
\ee 
and the Gauss-Codazzi equation 
\be 
\partial_{\chi}K 
   =  2 K_{ij} K^{ij} + q^{ij} \partial_{\chi} K_{ij} 
   = q^{ij} R_{ij} - {}^{(3)}\!R + K^2 
   =: - m^2 , 
\label{eq:GC} 
\ee 
we obtain 
\bea 
&&{}
   \left( \Box_3 - m^2 \right) \phi (\zeta^i) 
   + J 
   = 0, 
\label{Eom:left}
\eea 
where 
\bea 
    J := {}^{(3)}\!D^j h_{\chi j} 
            + \frac{1}{2} 
                         \partial_{\chi} h^j{}_j 
            - \left(
                    K + \frac{1}{2} \frac{\rho}{\sigma} 
              \right) h_{\chi \chi} . 
\label{Eq:J}
\eea 

Considering the perturbation of the wall motion in $M_-$, 
we obtain the equation for $\phi$ in the 
same form as Eqs. (\ref{Eom:left}) and (\ref{Eq:J}). 

Replacing the extrinsic quantities ( $K_{ij}, K^2, R_{ij}$) of $\Sigma$ 
by the averaged values ($\overline{K}_{ij}, \overline{K^2}, \overline{R}_{ij}$)
and using Eq.~(\ref{eq:WM}), we obtain 
\be 
 \left(
        \Box_3  - \overline{m^2}
 \right) {\phi (\zeta^i)}  
         + \overline{J} = 0, 
\label{Eom:3dim}
\ee 
with  
\bea 
  \overline{m^2} &=& - \overline{R}_{ij}q^{ij} + {}^{(3)}\!R 
                     - \left( \frac{\rho}{\sigma}\right)^2, 
\\ 
  \overline{J} 
          &=& {}^{(3)}\!D^j  h_{\chi j} 
              - \frac{1}{2} \overline{ \partial_{\chi} h^j{}_j } 
              + \frac{1}{2}
                           \left(
                                 \frac{\rho}{\sigma} 
                           \right) 
                                  h_{\chi \chi}. 
\eea  
This is the perturbed EOM for a domain wall coupled to 
gravitational perturbations. Note that the source term $J$ is 
described by the longitudinal part, the trace part, and the scalar part 
of the metric perturbations. In the derivation of Eq.~(\ref{Eom:3dim}), 
no symmetry is required to the background geometry. 

In relativistic perturbation theory, perturbed quantities have 
unphysical gauge freedom. 
Associated with an infinitesimal coordinate transformation 
\be 
\bar\del: 
         x^{\mu} \longrightarrow x^{\mu} + \bar\del x^{\mu} 
                                 =: x^{\mu} + \xi^{\mu}, 
\label{transf:infcoordinate}
\ee  
the metric perturbations transform as 
\be 
\bar{\del}: h_{\mu \nu} \longrightarrow 
     h_{\mu \nu} + \bar\del h_{\mu \nu} 
     := h_{\mu \nu} - {\nabla}_\mu {\xi}_\nu - {\nabla}_\nu {\xi}_\mu. 
\label{gaugetransfmetric}
\ee 
As seen below, though Eq.~(\ref{Eom:3dim}) is gauge-invariant, 
the scalar field $\phi$ and $J$ themselves are gauge-dependent. 
Indeed, associated with the transformation~(\ref{transf:infcoordinate}), 
$\phi$ transforms, by the definition~(\ref{scalar}), as 
\be
   \bar\del \phi = \xi_\chi, 
\label{gauge:scalar} 
\ee 
and $J$ transforms as 
\be 
   \bar\del J = - (\Box_3 - m^2) \xi_\chi - \xi^j D_j K , 
\label{del:J}
\ee 
where we use the equation 
\be 
   q^{ij} \partial_\chi \tD_i \xi_j 
           = \partial_\chi \tD^l \xi_l - 2 K^{ij} \tD_i \xi_j. 
\ee 
Once the average is taken, the second term in the RHS of Eq.~(\ref{del:J}) 
vanishes from Eq.~(\ref{eq:WM}). 
This suggests that the wall variable $\phi$ combines with 
a gauge-dependent gravitational variable contained in $J$ 
to form a gauge-invariant wall displacement variable. 
We shall represent Eq.~(\ref{Eom:3dim}) in terms of the gauge-invariant 
variables explicitly in general spherically symmetric background case 
in the next section.

%%%%%%%%%%%%%%%%%%%%%%%%%%%%%%%%%%%%%%%%%%%%%%
\subsection{Spherically symmetric background case}
%%%%%%%%%%%%%%%%%%%%%%%%%%%%%%%%%%%%%%%%%%%%%%

In spherically symmetric background case we can expand 
the perturbations by the tensor harmonics and express 
the perturbed equation in terms of the harmonic expansion coefficients. 
Furthermore, introducing the gauge-invariant variables for the metric 
perturbations according to Gerlach and Sengupta~\cite{GS3}, 
we find the gauge-invariant wall displacement variable. 
Then, we write down Eq.~(\ref{Eom:3dim}) explicitly 
in the gauge-invariant way.  

In general spherically symmetric spacetime, the metric takes 
the form 
\be
   ds^2  = g_{\mu\nu}dx^\mu dx^\nu  
         = \gamma_{ab}(y^c) dy^a dy^b + r^2(y^c) \Omega_{pq}dz^p dz^q. 
\ee
Here, $\Omega_{pq}dz^p dz^q$ is the metric on a unit symmetric two-sphere
with angular coordinates $z^p$, i.e.
\be
        \Omega_{pq}dz^p dz^q = d\theta^2 + \sin^2\theta d\varphi^2
                 =: d\Omega^2, 
\ee  
where $\theta, \varphi$ are the standard angular coordinates. 
The functions $r(y^c)$ and $\gamma_{ab}(y^c)$
are scalar and tensor fields on the two-dimensional
orbit space ${}^2M$ spanned by the two-coordinates $y^a$
and each point of ${}^2M$ represents a symmetric two-sphere. 
The indices $a,b,c,... $ represent indices on the orbit space 
and $p,q,r,...$ on the two-sphere. 
The history of the spherically symmetric domain wall is 
described by a timelike orbit in ${}^2\!M$. 

Let $\tau^{a}$ be the future directed unit timelike vector 
of the orbit of $\Sigma$ in ${}^2\!M$. The proper time $\tau$ of $\Sigma$ 
is defined by $\tau^a \partial_a \tau = 1$. 
Then, the orbit space metric can be decomposed as
\be
   \gamma_{ab}
              = - \tau_a \tau_b + n_a n_b, 
\ee 
where $n_a$ is the unit normal vector to $\Sigma$ in ${}^2\!M$. 
The wall intrinsic metric is expressed as 
\bea 
q_{ij} d\zeta^i d\zeta^j 
&=&{} 
     - \tau_a \tau_b dy^a dy^b + q_{pq} dz^p dz^q 
\non \\ 
&=&{} 
     - d{\tau}^2 + r^2({\tau}) d\Omega^2. 
\eea 
The intrinsic coordinates $\zeta^{i}$ are naturally taken by 
\be 
\zeta^{i} = ({\tau}, \theta, \varphi). 
\ee  

Let $D_a$ and $\hat{D}_p$ be the covariant derivatives with respect to the
orbit space metric~$\gamma_{ab}$ and the unit two-sphere metric $\Omega_{pq}$, 
respectively. 
Then, the components of the extrinsic curvature $K_{ij} d\zeta^i d\zeta^j$ 
are represented by  
\bea
        K_{ab} &=& - \tau_a \tau_b K^c{}_c  
                   = - \tau_a\tau_b \tau^c D_{\para}n_c, 
\label{Kab}
\\
        K_{ap} &=& 0, \\
        K_{pq} &=& - \frac{1}{2} K^r{}_r q_{pq}  
                = - r^2 \Omega_{pq} \frac{D_{\perp} r}{r}, 
\label{Kpq}   
\eea 
where and hereafter, we use the abbreviated notation such as 
$f_{\para} := \tau^a f_a$ and $f_{\perp} := n^a f_a$. 
We also note that $K^c{}_c = - \tau^{\mu} \tau^{\nu}K_{\mu \nu}$, 
the partial trace $K^p{}_p = K^2{}_2 + K^3{}_3$, $D_a n^a$, 
and $D_{\perp}r$ are scalar fields on ${}^2M$.   

% ----------------------------------------------------------
\subsubsection{\bf Tensor harmonics expansion} 
% ----------------------------------------------------------

We introduce the vector and tensor harmonics on the unit symmetric two-sphere, 
which are induced from the spherical harmonic functions $Y^m_l(z^p)$ 
satisfying 
\be 
\left\{
       \hat{\Delta}_2 + l(l + 1)
\right\} Y^m_l
        = 0, 
\ee 
where $\hat{\Delta}_2 := \Omega^{pq} \hat{D}_p \hat{D}_q $. 

%----------------------------------------------------------------------------
The vector and tensor harmonics are given by, 
for the odd mode,  
\bea
&&{} (V^{(o)}{}_{lm})_p := \epsilon_{pq}\hat{D}^q Y^m_l, 
\\
&&{} 
(T^{(o2)}{}_{lm})_{pq} 
     := \frac{1}{2}(\epsilon_{qr}\hat{D}_p + \epsilon_{pr}\hat{D}_q ) 
        \hat{D}^{r}Y^m_l, 
\eea 
and for the even mode, 
\bea 
&&{} (V^{(e)}{}^{m}_{l})_p 
      := \hat{D}_p Y^{m}_{l}, 
\\ 
&&{} (T^{(e0)}{}^{m}_{l})_{pq} 
      := \frac{1}{2} \Omega_{pq} Y^{m}_{l}, \\
&&{} (T^{(e2)}{}^{m}_{l})_{pq} 
      := \left\{
                 \hat{D}_p \hat{D}_q + \frac{1}{2}l(l+1) \Omega_{pq} 
         \right\}
                 Y^{m}_{l}. 
\eea 
It is obvious that 
\bea
&&{}
    \hat{D}_p (V^{(o)}{}_{lm})^{p} = 0, 
    \qquad (T^{(o2)}{}_{lm}){}_p{}^p = 0, 
\label{ttodd}
\\
&&{} 
(T^{(e2)}{}_{lm}){}_p{}^p = 0. 
\label{traceless}
\eea 
%----------------------------------------------------------------------------- 
Then, the odd modes are irrelevant for $J$ given by Eq.~(\ref{Eq:J}), 
which contains neither transverse nor traceless part of 
the metric perturbations. 

The even mode metric perturbations $h_{\mu \nu}$ are expanded 
as
\bea 
h_{\mu \nu}dx^{\mu}dx^{\nu}
&=& 
  \sum_{l,m} 
    \Bigl\{
          (f^{(e)}{}_{lm})_{ab}Y_{lm} 
                                   dy^{a}dy^{b} 
          + r(f^{(e1)}{}_{lm})_{a}(V^{(e)}{}_{lm})_{p}
                                  \left( dy^{a}dz^{p} + dz^{p}dy^{a} \right) 
\non \\
& & {} + \left( 
                 r^2 (f^{(e0)}{}_{lm}) (T^{(e0)}{}_{lm})_{pq} 
               + r^2 (f^{(e2)}{}_{lm}) (T^{(e2)}{}_{lm})_{pq} 
         \right) 
               dz^{p}dz^{q}
    \Bigr\}. 
\label{evenharmoexp}
\eea 
Hereafter, the angular integers $l,m$ and 
summation~$\sum_{l,m}$ are suppressed for simplicity. 
The expansion coefficients, $f(y^c), f_a(y^c), f_{ab}(y^c)$ 
are scalar, vector, and symmetric tensor fields on the orbit space ${}^2\!M$,
respectively. 

The scalar function $\phi(\zeta^i) = \phi(\tau) Y^m_l (z^p)$ 
and $J$ are described by the even mode functions. 
After straightforward manipulation, we can express Eq.~(\ref{Eom:left}) 
in the form
\bea 
&&{} 
  \left\{ 
          - D^2_{\para} - 2 \left( \frac{D_{\para} r}{r} \right) D_{\para} 
          - \frac{l(l + 1)}{r^2} - m^2 
  \right\}
          \phi({\tau}) 
\non \\ 
&&{} 
\qquad \qquad 
 - D_{\para} f^{(e)}_{\para \perp} 
 - 2 \left( \frac{D_{\para} r}{r} \right) f^{(e)}_{\para \perp}  
 + \frac{1}{2} D_{\perp} f^{(e)}_{\para \para}  
 - \frac{l(l+1)}{r} f^{(e1)}_{\perp} - \frac{1}{2} D_{\perp} f^{(e0)} 
 - \left( 
          K + \frac{1}{2} \frac{\rho}{\sigma} 
   \right) f^{(e)}_{\perp \perp} 
 = 0,
\label{Eom:spherical}
\eea 
where we omit $Y^m_l (z^p)$. 
The operator acting on $\phi$ in the first term 
is nothing but the Klein-Gordon operator 
on the background world sheet. 

%----------------------------------------------------------
\subsubsection{\bf Gauge-invariant expression} 
%----------------------------------------------------------
Let us introduce the gauge-invariant perturbation variables and 
express Eq.~(\ref{Eom:spherical}) in terms of them. 

The even mode generator $\xi_{\mu}dx^{\mu}$ of the gauge 
transformation~(\ref{transf:infcoordinate}) is expanded 
as
\bea
&&{}
\xi^{(e)}_{\mu}dx^{\mu} = \xi^{(e)}_{a} Y dy^{a}
                               + r \xi^{(e)} V^{(e)}_p dz^{p}.
\eea 
Then, the even mode gauge transformed metric perturbations are
\bea
 \bar\del h_{\mu \nu}dx^{\mu}dx^{\nu} 
&=& 
    - ( D_a \xi^{(e)}_b + D_b \xi^{(e)}_a ) Y dy^a dy^b 
    - \left\{ 
              \xi^{(e)}_a + r^2 D_a \left( \frac{\xi^{(e)}}{r^2} \right) 
      \right\} V^{(e)}_{p} (dy^a dz^p + dz^p dy^a) 
\non \\ 
&&{} 
\qquad
     + 2 \left\{ 
                l(l + 1) \xi^{(e)} - 2r \xi^{(e)}_a D^ar 
         \right\} T^{(e0)}_{pq} dz^p dz^q 
     - 2 \xi^{(e)} T^{(e2)}_{pq} dz^p dz^q . 
\label{trnsf:gauge} 
\eea  
From Eqs.~(\ref{evenharmoexp}) and~(\ref{trnsf:gauge}), 
we take the even mode gauge-invariant metric perturbation variables as 
\bea
  {\cal F} &:=& f^{(e0)} + l(l+1) f^{(e2)} - \frac{4}{r} X^a D_a r,  
\label{def:F} \\ 
  {\cal F}_{ab} &:=&  f_{ab} - D_a X_b - D_b X_a 
                     + \frac{1}{2} {\cal F} \gamma_{ab}, 
\label{def:Fab} 
\eea 
where the vector $X^{a}$ is defined by
\begin{equation}
X^{a} := r f^{(e1)a} - \frac{1}{2} r^2 D^{a} f^{(e2)}. 
\label{def:Xa} 
\end{equation}
In terms of ${\cal F}_{ab}$ and ${\cal F}$, $J$ is represented as 
\bea 
J 
&=& 
\biggl[ 
 - D_{\para} {\cal F}_{\para \perp} 
 - 2 \left( \frac{D_{\para} r}{r} \right) {\cal F}_{\para \perp}  
 + \frac{1}{2} D_{\perp} {\cal F}_{\para \para}  
 - \frac{1}{4} D_{\perp} {\cal F}  
 - \left( 
          K + \frac{1}{2} \frac{\rho}{\sigma} 
   \right) 
          \left( {\cal F}_{\perp \perp} - \frac{1}{2} {\cal F} \right) 
\non \\
\qquad \qquad 
 &&{}
  +   
   \left\{ 
          - D^2_{\para} - 2 \left( \frac{D_{\para} r}{r} \right) D_{\para} 
          - \frac{l(l + 1)}{r^2} +  D_\perp K 
  \right\} X_{\perp}
\non \\
\qquad \qquad 
 &&{}
 - \left(
         K + \frac{\rho}{\sigma}
   \right) D_{\perp} X_{\perp}
 - X_{\para} D_{\para}K 
\biggl] Y, 
\eea 
where we use $\tau_c D_{\perp}n^c = 0$, Eqs.~(\ref{Kab}), and (\ref{Kpq}). 
%----------------------------------------------------------------------------- 
From Eq.~(\ref{eq:WM}), the average of the last line vanishes, i.e.,  
\be 
   \left(
         \overline{K} + \frac{\rho}{\sigma}
   \right) \overline{ D_{\perp} X_{\perp}} 
 = \overline{ X_{\para} D_{\para} {K}}
 = 0. 
\ee 
Thus, taking the average we obtain consequently 
\bea 
&&{} 
 \left(  \Box_3 - \overline{m^2}  \right) 
         {\mit \Xi} + \overline{J} = 0, 
\label{Eom:sphrclGgInv}  
\eea 
where 
\bea
        {\mit \Xi} := \left( \phi + X_{\perp} \right) Y
\label{def:giwdv}
\eea
is the wall displacement variable and 
\bea
 \overline{J}
 = \biggl\{
 - \overline{ D_{\para} {\cal F}_{\para \perp} } 
 - 2 \left( \frac{D_{\para} r}{r} \right) \overline{{\cal F}_{\para \perp}}
 + \frac{1}{2} \overline{ D_{\perp} {\cal F}_{\para \para} } 
 - \frac{1}{4} \overline{ D_{\perp} {\cal F} }  
 + \frac{1}{2}
      \left( 
             \frac{\rho}{\sigma} 
      \right) 
             \left( 
                   \overline{{\cal F}_{\perp \perp}} 
                   - \frac{1}{2}\overline{{\cal F}}
             \right)  
\biggl\} Y
\eea
is the source term. 
Since, for the transformation~(\ref{transf:infcoordinate}), 
the vector $X^{a}$ transforms as 
\be
\bar{\del} X^{a} = -\xi^{a}, 
\label{eq:X} 
\ee 
and from Eq.~(\ref{gauge:scalar}), 
it turns out that the wall displacement variable ${\mit \Xi}$ given by 
Eq.~(\ref{def:giwdv}) is gauge-invariant. 
Thus, Eq.~(\ref{Eom:sphrclGgInv}) is a manifestly 
gauge-invariant form of Eq.~(\ref{Eom:3dim}) 
for general spherically symmetric background case.

In the gauge choice such that $X_{\perp} = 0$ on the perturbed world sheet, 
$\phi(\tau)$ represents the displacement of the perturbed wall 
from the background world sheet. 
On the other hand, in the gauge $\phi = 0$ on the perturbed world sheet, 
which is used in Refs.~\cite{KIF,II}, 
$X_{\perp}$ measures the wall displacement. 

In the case of a vacuum bubble nucleation in a false vacuum sea, 
provided that $M_{-}$ is Minkowski spacetime and $M_{+}$ 
de Sitter spacetime with the cosmological constant 
$\Lambda =: 3 H^{2} = \kappa \rho $, the boundary of the bubble is 
a domain wall whose world sheet has an O(3,1) symmetry. 
Then, using the constraints of the Einstein equations~(see Ref.~\cite{II} 
or~\cite{GS1}),  
\bea 
{\cal F}^a{}_a = {\cal F}, \qquad 
D_b {\cal F}^b{}_a  = D_a {\cal F}, 
\eea 
we can easily verify that Eq.~(\ref{Eom:sphrclGgInv}) reduces to 
Eq.~(5.3) in Ref.~\cite{II} in the gauge choice $\phi = 0$. 

For such an O(3,1) symmetric background case, Eq.~(\ref{Eom:3dim}) 
is also described in a simple way analogous to the usual 
cosmological perturbation~\cite{CP}. 
In Appendix~\ref{app:O(4)}, we introduce the gauge-invariant variables 
following Ref.~\cite{CP} and express the equation in terms of them.

\section{Summary}   
We showed the procedure to obtain the EOM for a domain wall coupled to 
gravitational field from the Nambu-Goto action.  
We should take care that the spacetime is singular at the world sheet of 
the wall, where the left and the right derivatives of the metric
(extrinsic quantities) are defined respectively and these do not coincide. 
First, we concentrated on each half spacetime and got two equations 
by the action principle for the world sheet as the boundaries 
of the half spacetimes. 
Next, in order to obtain the EOM which is consistent with 
the Einstein equations we should take the average of these equations, 
in other words, take the average of the left and the right derivatives.

Following the proposed procedure, 
we derived the perturbed EOM for the domain wall in general background 
geometries with taking the gravitational back reaction into account. 
The effects of the gravitational back reaction is included 
in the source term of the equation.

In treatment of gravitational perturbations, there is the gauge ambiguity. 
In spherically symmetric background case, as a simple example, 
we found that $\phi$ combines with a gravitational variable to form 
a gauge-invariant wall displacement variable, and derived the EOM in 
a manifestly gauge-invariant form. 

In the special cases considered in Refs.~\cite{KIF} and~\cite{II}, 
wall's EOMs, which are derived by the use of the metric junction conditions 
in these works, coincide with Eq.~(\ref{Eom:sphrclGgInv}) 
if we take the gauge $\phi=0$. 
Since the metric junction conditions are nothing but 
the Einstein equations at the world sheet, this coincidence is quite natural. 
Indeed, the Einstein equations contain the EOM for matter via
\be 
   \nabla_\nu T^{\mu \nu} = 0, 
\ee 
which is ensured by the Bianchi identity. 
In this viewpoint, we may expect that, once we regard the orbit of a 
domain wall as a singular hypersurface, 
the dynamics of the wall is completely described in terms of 
the dynamics of gravitational field as discussed by 
Einstein, Infeld, and Hoffmann~\cite{EIH}. 
In Refs.~\cite{KIF} and~\cite{II}, 
for the case of perturbations of the O(3,1) symmetric domain walls, 
it is shown that the walls lose their dynamical degrees 
of freedom. In addition, there is no spontaneous oscillation of the wall. 
These results suggest that the gravitational back reaction term $J$ in 
Eq.~(\ref{Eom:3dim}) is important for the motion of domain walls 
coupled to gravitational perturbations. 

In order to clarify the feature of the coupled system of extended objects 
and gravitation in more detail, we should analyze more generic cases. 
To find a gauge-invariant form of Eq.~(\ref{Eom:3dim}) in generic 
background case is an interesting problem. 
Generalization of the study for the system of a cosmic string and 
gravitational field is one of the next steps.

\begin{center}
{\large {\bf Acknowledgements}}
\end{center}
We are grateful to Dr. Kouji Nakamura and Professor Akio Hosoya for discussion 
and their continuous encouragement. 
We also would like to thank Dr. Takahiro Tanaka for useful comments.  

%----------------------------------------------------------------------------  
\appendix  
%----------------------------------------------------------------------------  

\section{Calculation of the connection term}
\label{app:cal}
%----------------------------------------------------------------------------- 

The second term in the LHS of Eq.~(\ref{eq:normaldir}) is decomposed as 
\bea 
&&{} 
n_{\mu} \del ({\mit \Gamma}^{\mu}{}_{\nu \lambda} 
                                q^{ij} e^{\nu}_{i}e^{\lambda}_{j})
  = n_{\mu} \del{\mit \Gamma}^{\mu}{}_{\nu \lambda} 
                                q^{ij} e^{\nu}_{i} e^{\lambda}_{j} 
  + n_{\mu} {\mit \Gamma}^{\mu}{}_{\nu \lambda} 
                                \del q^{ij}e^{\nu}_{i} e^{\lambda}_{j} 
  + 2n_{\mu} {\mit \Gamma}^{\mu}{}_{\nu \lambda} 
                                 q^{ij}e^{\nu}_{i} 
\del e^{\lambda}_{j}. 
\label{app:Gamma} 
\eea 
From Eq.~(\ref{del:g}), the first term in the RHS of Eq.~(\ref{app:Gamma}) 
is written as  
\bea  
 n_{\mu} \del{\mit \Gamma}^{\mu}{}_{\nu \lambda} 
                                 q^{ij} e^{\nu}_{i}e^{\lambda}_{j}
 &=&{} 
  \frac{1}{2}n^{\mu}q^{ij} 
  \left( 
        \nabla_i h_{\mu j} 
        + \nabla_{j} h_{\mu i} - \nabla_{\mu} h_{ij}
  \right) 
\non \\ 
& &{} 
  - n^{\mu}q^{ij}\left\{ 
                         \nabla_i (\phi K_{\mu j}) + \nabla_{j} 
                        (\phi K_{\mu i}) - \nabla_{\mu} (\phi K_{ij}) 
                 \right\} 
\non \\
&=&{} 
   {}^{(3)}\! D^j h_{\chi j} - \frac{1}{2} q^{ij} \partial_{\chi} h_{ij} 
    - K h_{\chi \chi} 
    + \phi q^{ij} \partial_{\chi} K_{ij} 
\label{35}
\eea 
The second term in the RHS of Eq.~(\ref{app:Gamma}) becomes 
\be
n_{\mu} {\mit \Gamma}^{\mu}{}_{\nu \lambda} 
                              \del q^{ij}e^{\nu}_{i} e^{\lambda}_{j} 
  = 2 \phi K_{ij} K^{ij} - K_{ij} h^{ij}, 
\ee 
and the last term of Eq.~(\ref{app:Gamma}) vanishes. 

%----------------------------------------------------------------------------- 
\section{Perturbed EOM for an O(3,1) symmetric domain wall} 
\label{app:O(4)}
%----------------------------------------------------------------------------- 

%%%
\def\htD{\check{D}}
%%% 

In a Gaussian normal coordinate system, the metric of a spacetime 
with an O(3,1) symmetric timelike hypersurface $\Sigma$ is described as 
\bea 
ds^2 
     &=& d\chi^2 + a(\chi)^2 \left( 
                                 -d\check\tau^2 + \cosh^2 \check\tau d\Omega^2 
                             \right),  
\eea            
where $\check{\tau}$ is the proper time of $\Sigma$ normalized by $a(0)$. 
Hereafter, we shall consider the three-metric $\check{q}_{ij}$ rescaled by 
$a^2$, i.e., $\check{q}_{ij}d\zeta^i d\zeta^j 
= -d\check\tau^2 + \cosh^2 \check\tau d\Omega^2$ and denote 
the covariant derivative with respect to $\check{q}_{ij}$ by $\htD_i$. 
Let the metric perturbations be 
\bea 
h_{\chi \chi} &=:& 2 \alpha, 
\\ 
h_{\chi j} &=:& a \beta_j, 
\\ 
h_{i j} &=:& 2 a^2 \left( 
                       h_L \check{q}_{ij} + h_{Tij}
                 \right),  
\eea 
where $\htD^l h_{Tlj} = 0$. 

Decompose the perturbations into the scalar, the vector, 
and the tensor type as follows: \\ 
for a vector $v^j$ in $\Sigma$, 
\be 
v^j = \htD^j v_L + v_T{}^j, 
\ee 
where 
\be 
\htD^l \htD_l v_L = \htD_l v^l,  
\qquad 
\htD_l v_{T}{}^l = 0, 
\ee 
for a second rank symmetric tensor $t_{ij}$ in $\Sigma$, 
\bea 
t_{ij} &=& \frac{1}{3} t_L \check{q}_{ij} 
               + \frac{1}{2}\left( 
                                  \htD_j t_{Ti} + \htD_i t_{Tj} 
                            \right)  
               + \left(  
                      \htD_i \htD_j - \frac{1}{3} \check{q}_{ij} 
                                      \htD^l \htD_l
                \right) t_T 
               + t_{TTij}  , 
\eea 
where 
\bea 
    t_L := t^l{}_l, \qquad 
    \htD_l t_T{}^l = 0, \qquad 
    \htD_l t_{TT}{}^l{}_j = 0, \qquad 
    t_{TT}{}^l{}_l = 0. 
\eea 

Associated with the infinitesimal coordinate 
transformation~(\ref{transf:infcoordinate}), 
the metric perturbations transform as 
\bea 
\mbox{tensor type}, \qquad &&{}  
\non 
\\ 
&&{}
   \bar\del h_{TTij} = 0, 
\\  
\mbox{vector type}, \qquad &&{}  
\non 
\\ 
&&{} 
   \bar\del {\beta}_{Tj} = - a \dot{\xi}_{Tj}, 
\\ 
&&{}
   \bar\del {h}_{Tj} = -  \xi_{Tj}, 
\\ 
\mbox{scalar type}, \qquad &&{}  
\non  
\\ 
&&{}  
   \bar\del {\alpha} = - \dot{\xi}_\perp, 
\\ 
&&{} 
   \bar\del {\beta}_L = - a \dot{\xi}_L - \frac{1}{a} \xi_\perp, 
\\ 
&&{} 
    \bar\del {h}_L = - \frac{1}{3} \htD^l \htD_l \xi_L 
                     - \frac{\dot{a}}{a} \xi_\perp,
\\  
&&{} 
   \bar\del {h}_T = - \xi_L, 
\eea 
where the {\em dot} denotes the $\chi$ derivative and 
$\xi_\perp := \xi^\chi$. 
Then, the gauge-invariant perturbation variables are chosen as 
\bea 
\mbox{tensor type}, \qquad &&{}  \non 
\\ 
&&{} 
h_{TTij}, 
\\ 
\mbox{vector type}, \qquad &&{}  \non 
\\ 
&&{} 
\sigma^j_{T} := a \beta^j_T - a^2 \dot{h}^j_T, 
\\ 
\mbox{scalar type}, \qquad &&{}  \non 
\\ 
&&{} 
{\mit \Phi} := {\cal R} - \frac{\dot{a}}{a} \sigma_{L}, 
\\ 
&&{} 
{\mit \Psi} := \alpha - \dot{\sigma}_{L}. 
\eea 
Here 
\be 
{\cal R} := h_L - \frac{1}{3} \htD^l \htD_l h_T, 
\qquad 
\sigma_{L} := a \beta_L - a^2 \dot{h}_T, 
\ee 
and their gauge transformations are  
\be
    \bar\del {\cal R} = - \frac{\dot{a}}{a} \xi_\perp, 
\qquad 
    \bar\del \sigma_{L} = - \xi_\perp. 
\ee     

In terms of these variables, Eq.~(\ref{Eom:3dim}) is rewritten as 
\be 
\left( 
      \Box_3 - \overline{m^2}  
\right) {\mit \Xi}
         - \frac{3}{2} \overline{ \dot{\mit \Phi}} 
         + \frac{\rho}{\sigma} \overline{{\mit \Psi}} 
         = 0, 
\ee 
where ${\mit \Xi} := \phi(\zeta^i) + \sigma_{L}$ is the gauge-invariant wall 
displacement variable.

\newpage
%%%%%%%%%%%%%%%%%%%%%%%%%%%%%%%%%%%%%%%

\end{document}